\documentclass{sig-alternate-2013}

\usepackage{url}
\usepackage{comment}
\usepackage{amssymb}
\usepackage{amsmath}
\usepackage{graphicx}

\usepackage[utf8]{inputenc}
\usepackage[T2A,T1]{fontenc}
\usepackage[french,russian,polish,english]{babel}
\usepackage{CJKutf8} %\usepackage{CJK}
\newfont{\mycrnotice}{ptmr8t at 7pt}
\newfont{\myconfname}{ptmri8t at 7pt}
\let\crnotice\mycrnotice%
\let\confname\myconfname%

\permission{}
\conferenceinfo{To Appear at MM'15,}{October 26--30, 2015, Brisbane, Australia.}
\copyrightetc{\copyright~2015}
\crdata{978-1-4503-3459-4/15/10.\\
DOI: http://dx.doi.org/10.1145/2733373.2806246}

\begin{document}

\title{Visual Affect Around the World:\\ A Large-scale Multilingual Visual Sentiment Ontology}
%\subtitle{}

\numberofauthors{6}
\author{
% 1st. author
\alignauthor Brendan Jou\titlenote{Denotes equal contribution.}\\
       \affaddr{Columbia University}\\
       \affaddr{New York, NY USA}\\
       \email{bjou@ee.columbia.edu}
% 2nd. author
\alignauthor Tao Chen\raisebox{9pt}{$\ast$}\\
       \affaddr{Columbia University}\\
       \affaddr{New York, NY USA}\\
       \email{taochen@ee.columbia.edu}
% 3rd. author
\alignauthor Nikolaos Pappas\raisebox{9pt}{$\ast$}\\
       \affaddr{Idiap Research Institute/EPFL}\\
       \affaddr{Martigny, Switzerland}\\
       \vspace{-.42mm}\email{npappas@idiap.ch}
\and
% 4th. author
\alignauthor Miriam Redi\raisebox{9pt}{$\ast$}\\
       \affaddr{Yahoo Labs}\\
       \affaddr{London, UK}\\
       \email{redi@yahoo-inc.com}
% 5th. author
\alignauthor Mercan Topkara\raisebox{9pt}{$\ast$}\\
       \affaddr{JW Player}\\
       \affaddr{New York, NY USA}\\
       \email{mercan@jwplayer.com}
% 6th. author
\alignauthor Shih-Fu Chang\\
       \affaddr{Columbia University}\\
       \affaddr{New York, NY USA}\\
       \email{sfchang@ee.columbia.edu}
}

\date{}

\maketitle

\begin{abstract}
Every culture and language is unique.
Our work expressly focuses on the uniqueness of culture and language in relation to human affect, specifically sentiment and emotion semantics, and how they manifest in social multimedia.
We develop sets of sentiment- and emotion-polarized visual concepts by adapting semantic structures called adjective-noun pairs, originally introduced by Borth et al.~\cite{borth_2013_vso}, but in a multilingual context.
We propose a new language-dependent method for automatic discovery of these adjective-noun constructs.
We show how this pipeline can be applied on a social multimedia platform for the creation of a large-scale multilingual visual sentiment concept ontology (MVSO).
Unlike the flat structure in \cite{borth_2013_vso}, our unified ontology is organized hierarchically by multilingual clusters of visually detectable nouns and subclusters of emotionally biased versions of these nouns.
In addition, we present an image-based prediction task to show how generalizable language-specific models are in a multilingual context.
A new, publicly available dataset of >15.6K sentiment-biased visual concepts across 12 languages with language-specific detector banks, >7.36M images and their metadata is also released.
\end{abstract}

% ACM CCS
%\category{H.1.2}{Models and Principles}{User/Machine Systems}
%\category{H.5.1}{Information Interfaces and Presentation}{Multimedia Information Systems}
\category{H.5.4}{Information Interfaces and Presentation}{Hypertext/Hypermedia}
\category{I.2.10}{Artificial Intelligence}{Vision and Scene Understanding}

%\terms{Emotion; Sentiment; Ontology; Multilingual}

\keywords{Multilingual; Language; Cultures; Cross-cultural; Emotion; Sentiment; Ontology; Concept Detection; Social Multimedia}

%============================================================
\section{Introduction} \label{sec:intro}

If you scoured the world and took several people at random from major countries and asked them to fill in the blank ``\_\_\_\_\_\_\_ love'' in their native tongue, how many unique adjectives would you expect to find?
Would people from some cultures tend to fill it with \emph{twisted}, while others \emph{pure} or \emph{unconditional} or \emph{false}?
All over the world, we daily express our thoughts and feelings in culturally isolated contexts;
and when we travel abroad, we know that to cross a physical border also means to cross into the unique behaviors and interactions of that people group -- its cultural border.
How similar or different are our sentiments and feelings from this other culture?
Or the thoughts and objects we tend to talk about most?
Motivated by questions like this, our work explores the computational understanding of human affect along cultural lines, with focus on visual content.
In particular, we seek to answer the following important questions:
(1) how are images in various languages used to express affective visual concepts, e.g.~\emph{beautiful place} or \emph{delicious food}? % image-to-ANPs
And (2) how are such affective visual concepts used to convey different emotions and sentiment across languages? % ANP-to-emotion/sentiment

\begin{figure}[t]
  \centering
  \includegraphics[scale=0.165]{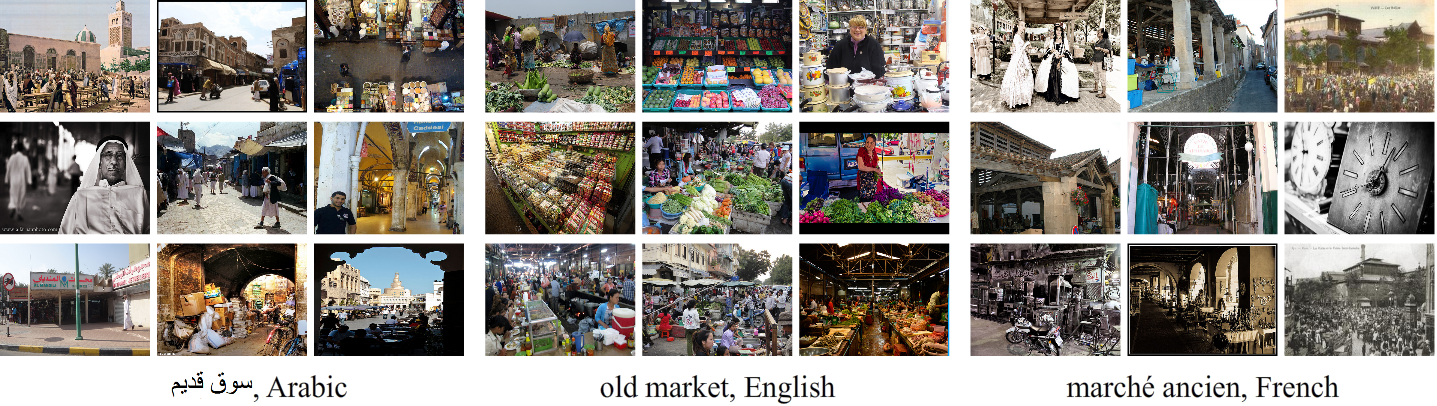}\\\vspace{2mm}
  \includegraphics[scale=0.165]{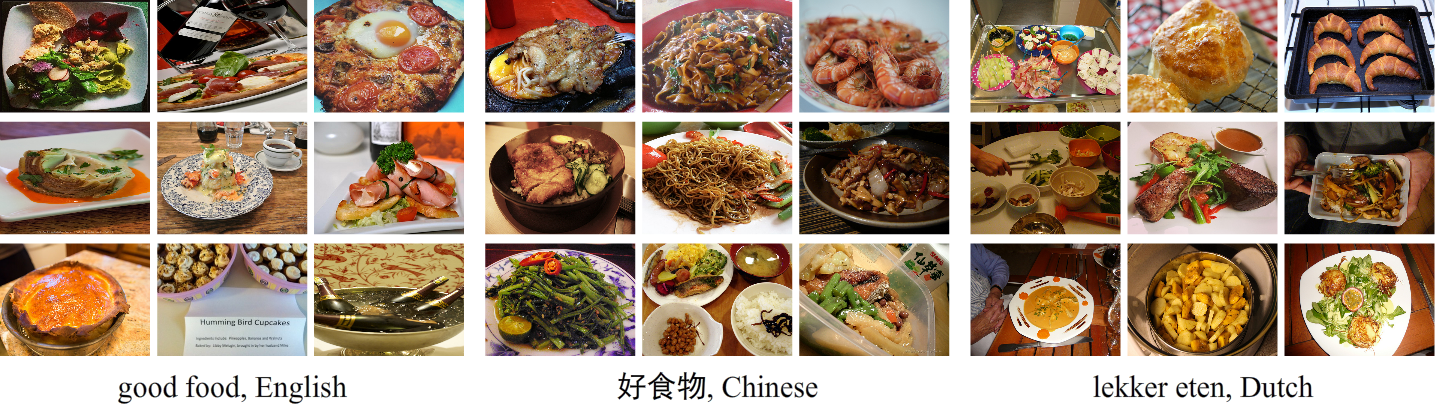}
  \vspace{-6mm}
  \caption{Example images from ``around the world'' organized by affective visual concepts. Top set shows images of \emph{old market} concept from three different cultures/languages; and the bottom, images of \emph{good food}. Even though conceptual reference is the same, each culture's sentimental expression of these concept may be adversely different.}
  \label{fig:teaser}
\end{figure}

\begin{figure*}[t]
  \centering
  \includegraphics[scale=0.62]{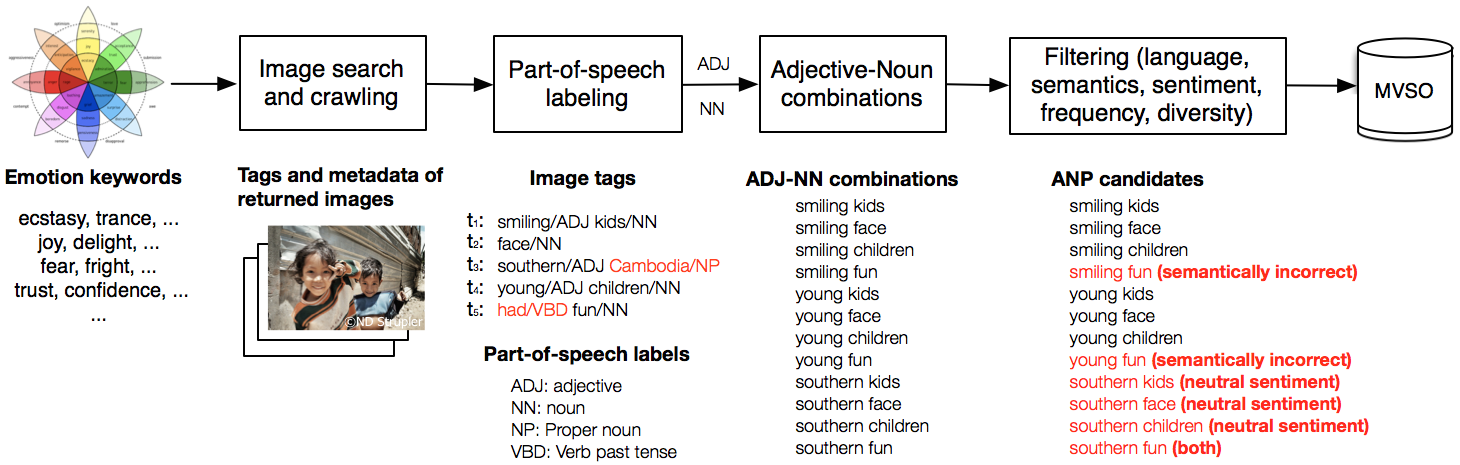}
  \caption{The construction process of our multilingual visual sentiment ontology (MVSO) begins with crawling images and metadata based on emotion keywords. Image tags ($t_{1}, \ldots, t_{5}$) are labeled with part-of-speech tags, and adjectives and nouns are used to form candidate adjective-noun pair (ANP) combinations \cite{borth_2013_vso}, while others are ignored (in red). Finally, these candidate ANPs are filtered based on various criteria (Sec.~\ref{sec:filtering}) which help remove incorrect pairs (in red), forming a final MVSO with diversity and coverage.}
  \label{fig:overview}
\end{figure*}

In Psychology, there are two major schools-of-thought on the connection between cultural context and human affect, i.e.~our experiential feelings via our sentiments and emotions.
Some believe emotion to be culture-specific \cite{mccarthy_1994}, that is, emotion is dependent on one's cultural context, while others believe emotion to be universal \cite{haselton_2006}, that is, emotion and culture are independent mechanisms.
For example, while this paper is written in English, there are emotion words/phrases in other languages for which there is no exact translation in English, e.g., \emph{Schadenfreude} in German refers to pleasure at someone else's expense.
Do English-speakers not feel those same emotions or do they simply refer to them in a different way?
Or even if the reference is the same, perhaps the underlying emotion is different?

In Affective Computing \cite{picard_1997} and Multimedia, we often refer to the \emph{affective gap} as the conceptual divide between the low-level visual stimuli, like images and features, and the high-level, abstracted semantics of human affect, e.g.~\emph{happy} or \emph{sad}.
In one attempt to bridge sentiment and visual media, Borth et al.~\cite{borth_2013_vso} developed a \emph{visual sentiment ontology} (VSO), a set of 1,200 mid-level concepts using structured semantics called adjective-noun pairs (ANPs).
The noun portion of the ANP allows for computer vision detectability and the adjective serves to polarize the noun toward a positive or negative sentiment, or emotion, e.g.~so instead of having visual concepts like \emph{sky} or \emph{dog}, we have \emph{beautiful sky} or \emph{scary dog}.
Many works like this that have built algorithms, models and datasets on the assumption of the psychology theory that emotions are universal.
However, while such works provide great research contributions in that native language, their applicability and generalization to other languages and cultures remains largely unexplored.

We present a large-scale multilingual visual sentiment ontology (MVSO) and dataset including adjective-noun pairs from 12 languages of diverse origins: Arabic, Chinese, Dutch, English, French, German, Italian, Persian, Polish, Russian, Spanish, and Turkish.
We make the following contributions:
\textbf{(1)} a principled, context-aware pipeline for designing a multilingual visual sentiment ontology,
\textbf{(2)} a Multilingual Visual Sentiment Ontology mined from social multimedia data end-to-end,
\textbf{(3)} a MVSO organized hierarchically into noun-based clusters and sentiment-biased adjective-noun pair subclusters,
\textbf{(4)} a multilingual, sentiment-driven visual concept detector bank, and
\textbf{(5)} the release of a dataset containing MVSO and large-scale image collection with benchmark cross-lingual sentiment prediction\footnote{\url{mvso.cs.columbia.edu}}.

\begin{table*}[t]
  \scriptsize
  \centering
  \begin{tabular}{c|c|c|c|c|c|c|c|c}
    \textbf{English} & joy & trust & fear & surprise & sadness & disgust & anger & anticipation \\
    \textbf{Spanish} & alegría & confianza & miedo & sorpresa & tristeza & asco & ira & previsión \\
    \textbf{Italian} & gioia & fiducia & paura & sorpresa & tristezza & disgusto & rabbia & anticipazione \\
    \textbf{French} & bonheur & confiance & peur & surprise & tristesse & \foreignlanguage{french}{dégoût} & colère & prévision \\
    \textbf{German} & Freude & Vertrauen & Angst & Überraschung & Traurigkeit & Empörung & Ärger & Vorfreude \\
    \textbf{Chinese} & \begin{CJK}{UTF8}{bkai}歡樂\end{CJK} & \begin{CJK}{UTF8}{bkai}信任\end{CJK} & \begin{CJK}{UTF8}{bkai}害怕\end{CJK} & \begin{CJK}{UTF8}{bkai}震驚\end{CJK} & \begin{CJK}{UTF8}{bkai}悲\end{CJK} & \begin{CJK}{UTF8}{bkai}討厭\end{CJK} & \begin{CJK}{UTF8}{bkai}憤怒\end{CJK} & \begin{CJK}{UTF8}{bkai}預期\end{CJK} \\
    \textbf{Dutch} & vreugde & vertrouwen & angst & verrassing & verdriet & walging & woede & anticipatie
  \end{tabular}
  \caption{Most representative keywords according to native/proficient speakers for eight basic emotions and for 7 of our 12 languages, chosen and shown top-to-bottom in decreasing no.\ of discovered visual affect concepts, or adjective-noun pairs.}
  \label{tab:kw_examples}
\end{table*}

%============================================================
\section{Related Work} \label{sec:related work}
We address the general challenge of \emph{affective image understanding}, aiming at both recognition and analysis of sentiment and emotions in visual data, but from a multilingual and cross-cultural perspective.
Our work is closely related to Multimedia and Vision research that focus on visual aesthetics \cite{ke_2006}, interestingness \cite{gygli_2013}, popularity \cite{khosla_2014}, and creativity \cite{redi_2014}.
Our work also relates to research in Cognitive and Social Psychology, especially emotion and culture research \cite{ekman_1993,mesquita_1997,plutchik_1980}, but also neuroaesthetics \cite{vessel_2014}, visual preference \cite{vessel_2014,zajonc_1980}, and social interaction \cite{markus_1991}.

Progressive research in ``visual affect'' recognition was done in \cite{yanulevskaya_2008} and \cite{machajdik_2010} where image features were designed based on art and psychology principles for emotion prediction.
And such works were later improved in \cite{jia_2012} by adding social media data in semi-supervised frameworks.
From this research effort in visual affect understanding, several affective image datasets were released to the public.
The International Affective Picture System (IAPS) dataset \cite{lang_1997} is a seminal dataset of $\sim$1,000 images, focused on induced emotions in humans for biometric measurement.
The Geneva Affective PicturE Database (GAPED) \cite{danglauser_2011} consists of 730 pictures meant to supplement IAPS and tries to narrow the themes across images.
And recently, in \cite{borth_2013_vso}, a \emph{visual sentiment ontology} (VSO) and dataset was created from Flickr image data, resulting in a collection of adjective-noun pairs along with corresponding images, tags and sentiment.
One major issue with these datasets and existing methods is that they do not consider the \emph{context} in which emotions are felt and perceived.
Instead, they assume that visual affect is universal, and do not account for the influence of culture or language.
We explicitly tackle visual affect understanding from a multi-cultural, multilingual perspective.
In addition, while existing works often use handpicked data, we gather our data ``in the wild'' on a popular, multilingual social multimedia platform.

The study of emotions across language and culture has long been a topic of research in Psychology.
A main contention in the area concerns whether emotions are culture-specific \cite{mccarthy_1994}, i.e.~their perception and elicitation varies with the context, or universal \cite{haselton_2006}.
In \cite{russell_1991}, a survey of cross-cultural work on semantics surrounding emotion elicitation and perception is given, showing that there are still competing views as to whether emotion is pan-cultural, culture-specific, or some hybrid of both.
Inspired by research in this domain, we are the first to investigate the relationship between visual affect and culture\footnote{Note that we use \emph{language} and \emph{culture} interchangeably often. We define language as the ``lens'' through which we can observe culture. So while the two can be distinguished, for simplicity, we use them interchangeably.} from a multimedia-driven and computational perspective, as far as we know.

Other work in cross-lingual research comes from text sentiment analysis and music information retrieval.
In \cite{bautin_2008} and \cite{mihalcea_2007}, they developed multilingual methods for international text sentiment analysis in online blogs and news articles, respectively.
In \cite{lee_2005} and \cite{hu_2014}, they presented approaches to indexing digital music libraries with music from multiple languages.
Specific to emotion, \cite{hu_2014} tried to highlight differences between languages by building models for predicting the musical mood and then cross-predicting in other languages.
Unlike these works, we propose a multimedia-driven approach for cross-cultural visual sentiment analysis in the context of online image collections.

It is important to distinguish our work from that of Borth et al.~on VSO \cite{borth_2013_vso} and its associated detector bank, SentiBank \cite{borth_2013_sb}.
Their mid-level representation approach has recently proven effective in a wide range of applications in emotion prediction \cite{borth_2013_sb,jou_2014}, social media commenting \cite{chen_2014}, etc.
However, in addition to lack of multilingual support, there are several technical challenges with VSO \cite{borth_2013_sb,borth_2013_vso} that we seek to improve on via
(1) detection of adjectives and nouns with language-specific part-of-speech taggers, as opposed to a fixed list of adjectives and nouns,
(2) automatic discovery of adjective-noun pairs correlated with emotions, as opposed to ``constructed'' pairs from top frequent adjectives and nouns, and
(3) stronger selection criterion based on image tag frequency, linguistic and semantic filters and crowdsource validation..
Our proposed MVSO discovery method can be easily extended to any language, while achieving greater coverage and diversity than VSO.

%============================================================
\section{Ontology Construction} \label{sec:framework}
An overview of the proposed method for multilingual visual sentiment concept ontology construction is shown in Figure \ref{fig:overview}.
In the first stage, we obtain a set of images and their tags using seed emotion keyword queries, selected according to emotion ontologies from psychology such as \cite{plutchik_1980} or \cite{ekman_1993}.
Next, each image tag is labeled automatically by a language-specific part-of-speech tagger and adjective-noun combinations are discovered from words in the tags.
Then, the combinations are filtered based on language, semantics, sentiment, frequency and diversity filters to ensure that the final set of ANPs have the following properties:
(a) are written in the target language,
(b) they do not refer to named entities or technical terms,
(c) reflect a non-neutral sentiment,
%(d) have a link to an emotion
(d) are frequently used, and
(e) are used by a non-trivial number of users of the target language.

The discovery of affective visual concepts for these languages using adjective-noun pairs poses several challenges in lexical, structural and semantic ambiguities, which are well-known problems in natural language processing.
Lexical ambiguity is when a word has multiple meanings which depend on the context, e.g.~\emph{sport jaguar} or \emph{forest jaguar}.
Structural ambiguity is when a word might have different grammatical interpretation depending on the position in the context, e.g.~\emph{ambient light} or \emph{light room}.
Semantic ambiguity is when a combination of words with the same syntactic structure have different semantic interpretation, e.g.~\emph{big apple}.
We selected languages in our MVSO according to the availability of public natural language processing tools and sentiment ontologies per language so that automatic processing was feasible.
In addition, we sought to cover a wide range of geographic regions from the Americas to Europe and to Asia.
We settled on 12 languages: Arabic, Chinese, Dutch, English, French, German, Italian, Persian, Polish, Russian, Spanish, and Turkish.

We applied our proposed data collection pipeline to a popular social multimedia sharing platform, Yahoo! Flickr\footnote{\url{www.flickr.com}}, and collected public data from November 2014 to February 2015 using the Flickr API.
We selected Flickr because there is an existing body of multimedia research using it in the past, and in particular, \cite{jin_2010} describes how Flickr satisfies two conditions for making use of the ``wisdom of the social multimedia'': popularity and availability. 
We do not repeat the argument in \cite{jin_2010}, but note that in addition to those benefits, Flickr has multilingual support and the use of Flickr facilitates a natural comparison to the seminal VSO \cite{borth_2013_vso} work.

\begin{table}[t]
  \scriptsize
  \centering
  \begin{tabular}{l|c|c|c|c}
     & \textbf{\#images} & \textbf{\#tags} & \textbf{\#cand} & \textbf{\#anps (final)}\\ \hline
    Arabic  &   116,125 &    958,435 &    15,532 &    29 \\
    Chinese &   895,398 &  3,919,161 &    50,459 &   504 \\
    Dutch   &   260,093 &  4,929,581 & 1,045,290 &   348 \\
    English & 1,082,760 & 26,266,484 & 2,073,839 & 4,421 \\
    French  &   866,166 & 22,713,978 & 1,515,607 & 2,349 \\
    German  &   528,454 & 10,525,403 &   854,100 &   804 \\
    Italian &   548,134 & 10,425,139 & 1,324,076 & 3,349 \\
    Persian &   128,546 &  1,304,613 &   103,609 &    15 \\
    Polish  &   294,821 &  5,261,940 &   141,889 &    70 \\
    Russian &    60,108 &  1,518,882 &    30,593 &   129 \\
    Spanish &   827,396 & 15,241,679 &   925,975 & 3,381 \\
    Turkish &   332,609 &  4,717,389 &    73,797 &   231 \\ \hline
    \textbf{\#total} & 5,940,610 & 107,782,684 & 8,154,766 & 15,630 \\
  \end{tabular}
  \caption{Ontology refinement statistics over 12 languages.
Beginning with many images from seed emotion keywords denoted by \#images, we extracted tags from these images \#tags, and performed adjective-noun pair (ANP) discovery for candidate combinations \#cand.
Through a series of filters -- frequency, language, semantics filter, sentiment filter and diversity -- and after crowdsourcing, we got our final visual sentiment concepts \#anps.}
  \label{tab:statistics}
\end{table}

\subsection{Adjective-Noun Pair Discovery} \label{sec:anp_discovery}
As our seed emotion ontology, we selected the \textit{Plutchik's Wheel of Emotions} \cite{plutchik_1980}.
This psychology ontology was selected because it consists of graded intensities for multiple basic emotions providing a richer set of emotional valences compared to alternatives like \cite{ekman_1993}; it has also been shown to be useful for VSO \cite{borth_2013_vso}.
The Plutchik emotions are organized by eight basic emotions, each with three valences:
ecstasy > joy > serenity;
admiration > trust > acceptance;
terror > fear > apprehension;
amazement > surprise > distraction;
grief > sadness > pensiveness;
loathing > disgust > boredom;
rage > anger > annoyance; and,
vigilance > anticipation > interest.

\textbf{Multilingual Query Construction}:
To obtain seeds for each language, we recruited 12 native and proficient language speakers to provide a set of translated or synonymous keywords to those of the 24 Plutchik emotions.
Speakers were allowed to use any number of keywords per emotion since the possible synonyms per emotion and language can vary, but they were asked to rank their chosen keywords along each emotion seed.
They were also allowed to use tools like Google Translate\footnote{\url{translate.google.com}} or other resources to enrich their emotion keywords.
Table \ref{tab:kw_examples} lists top ranked keywords according to speakers for 7 out of 12 languages in each emotion.

Given the set of keywords $E^{(l)}=\{e_{ij}^{(l)}\ |\ i=1\ldots 24, j=1\ldots n_{i} \}$ describing each emotion $i$ per language $l$, where $n_{i}$ is the number of keywords per emotion $i$, we performed tag-based queries on tags with the Flickr API to retrieve images and their related tags.
Like \cite{borth_2013_vso}, for each emotion, we chose to sample only the top 50K images ranked by Flickr relevance to simply limit the size of our results, but if an emotion had less than 50K images, we extended the search to additional metadata, i.e.~title and description.

\textbf{Part-of-speech Labeling}:
To identify the type of each word in a Flickr tag, we performed automatic part-of-speech labeling using pre-trained language-specific taggers which achieve high accuracy (>95\% for most languages), namely TreeTagger \cite{schmid_1994}, Stanford tagger \cite{toutanova_2003}, HunPos tagger \cite{halcasy_2007} and a morphological analyzer for Turkish \cite{zelal_1994}.
Though not all the tags contained multiple words, the average number of words was always greater than the average number of tags for all languages, so word context was almost always taken into account.
From the full set of part-of-speech labels, we retained identified nouns, adjectives and other part-of-speech types which can be used as adjectives, such as simple or past participle (e.g.~\emph{smiling face}) in English.

\textbf{Discovery Strategy}: 
We based our discovery strategy for ANPs on co-occurrence in image tags, that is, if an adjective-noun pair is relevant to the specific emotion it should appear at least once as that exact pair phrase in the crawled images for that emotion.
To validate the completeness of our strategy we compared with VSO and found that $\sim$86\% of ANPs discovered by VSO \cite{borth_2013_vso} overlap with the English ANPs discovered by our method.

\subsection{Filtering Candidate Adjective-Noun Pairs} \label{sec:filtering}
From these discovered ANPs, we applied several filters to ensure they satisfied the following criteria:
(a) written in the target language,
(b) do not refer to named entities,
(c) reflect a non-neutral sentiment,
(d) frequently used and
(e) used by multiple speakers of the language.

\textbf{Language \& Semantics}:
We used a combination of language dictionaries\footnote{\url{www.winedt.org}} instead of language classifiers to verify the correctness of the ANP as the performance of using the latter was low for short-length text, especially for Romance languages which share characters. 
All of the English ANPs were classified as indeed English by the dictionary, while for other languages, ANPs were removed if they passed the English dictionary filter but not the target language dictionary.
The intuition for this was that most candidate ANPs in other languages were mixed mostly with English.
We removed candidate pairs which referred to named entities or technical terms, where named entities were detected using several public knowledge bases such as Wikipedia and dictionaries for names\footnote{\url{www.wikipedia.org} and \url{www.ssa.gov}, respectively}, cities, regions and countries\footnote{\url{www.geobytes.com}}, and technical terms were removed via a manually created list of words specific to our source domain, Flickr, containing photography-related (e.g.~\emph{macro}, \emph{exposure}) and camera-related words (e.g.~\emph{DSLR}).

\textbf{Non-neutral Sentiment}:
To filter out neutral candidate adjective-noun pairs, each ANP was scored in sentiment using two publicly available sentiment ontologies: SentiStrength \cite{thelwall_2010} and SentiWordnet \cite{esuli_2006}.
SentiStrength ontology supported all the languages we considered, but since SentiWordnet could only be used directly for English, we passed in automatic translations in English from all other languages to it, following previous research on multilingual sentiment analysis in machine translation \cite{balahur_2012,carmen_2008}\footnote{For four non-English languages with the highest ANP counts, we have verified only a small percentage of non-neutral ANPs (less than 2\%) reverse sentiment polarity after translation, confirming similar observations in the previous work.}.
We computed the ANP sentiment score $S(anp) \in [-2,+2]$ as: 
\begin{equation}
  S(anp) = \left\{
     \begin{array}{ll}
         S(a) \hspace{-1mm} &: sgn\left\{S(a)\right\} \neq sgn\left\{S(n)\right\} \\
         S(a) + S(n) \hspace{-1mm} &: \textrm{otherwise}
     \end{array}
    \right.
  \label{eq:sent}
\end{equation}
where $S(a) \in [-1,+1]$ and $S(n) \in [-1,+1]$ are the sentiment scores of the individual adjective and noun words, respectively, each of which are given by the arithmetic mean of SentiStrength and SentiWordnet scores on the word, and $sgn$ is the sign of the scores.
The piecewise condition essentially says that if the signs of the sentiment scores of the adjective and noun differ, then we ignore the noun.
This highlights our belief that adjectives are the dominant sentiment modifiers in an adjective-noun pair, so for example, even if a noun is positive, like \emph{wedding}, an adjective such as \emph{horrible} would completely change the sentiment of the combined pair.
And so, for these sign mismatch cases, we chose the adjective's sentiment alone.
In the other case, when the sign of the adjective and noun were the same, whether both positive (e.g.~\emph{happy wedding}) or both negative (e.g.~\emph{scary spider}), we simply allowed the ANP sentiment score to be the unweighted sum of its parts.
ANP candidates with zero sentiment score were filtered out.

\textbf{Frequency}:
Good ANPs are those which are actually used together.
Here, we loosely defined an ANP's ``frequency'' of usage as its number of occurrences as an image tag on Flickr.
When computing counts for each pair, we accounted for language-specific syntax like the ordering of adjectives and nouns.
Following anthropology research \cite{wals}, we followed two dominant orderings (91.5\% of the languages worldwide): adj-noun and noun-adj.
We also ``merged'' simplified and traditional forms in Chinese by considering them to be from the same language pool but distinct characters sets.
In addition, we considered the possible intermediate Chinese character \begin{CJK}{UTF8}{bkai}的\end{CJK} during our frequency counting.
For all non-English languages, we retained all ANPs that occurred at least once as an image tag; but for English, since Flickr's most dominant number of users are English-speaking, we set a higher frequency threshold of 40. 

\textbf{Diversity}:
The shear frequency of an adjective-noun pair occurrence alone was not sufficient to ensure a pair's pervasive use in a language.
We also checked if the ANP was used by a non-trivial number of distinct Flickr users for a given language.
We identified the number of users contributing to uploads of images for each ANP and found a power law distribution in every language.
To avoid this uploader bias, we removed all ANPs with less than three uploaders.
Many removed candidate pairs came from companies and merchants for advertising and branding.

To further ensure diversity in our MVSO, we subsampled nouns in every language by limiting to the 100 most frequent ANPs per adjective so that we do not have, for example, the adjective \emph{surprising} modifying every possible noun in our corpus.
In addition, we performed stem unification by checking and including only the inflected form (e.g.~singular/plural) of an ANP that was most popular in usage as a tag on Flickr.
This unification did also filter some candidate ANPs as some ``duplicates'' were present but simply in different inflected forms.

\subsection{Crowdsourcing Validation} \label{sec:crowdsourcing}
A further inspection of the corpus after the automatic filtering process showed that some issues could not completely be solved in an automatic fashion.
Common errors included many fundamental natural language processing challenges like confusions in named entity recognition (e.g.~\emph{big apple}), language mixing (e.g.~adjective in English + noun in Turkish), grammar inconsistency (e.g.~adj-adj, or verb-noun) and semantic incongruity (e.g.~\emph{happy happiness}).
So to refine our multilingual visual sentiment ontology, we crowdsourced a validation task.
For each language, we asked native speaking workers to evaluate the correctness of ANPs post automatic filtering.
We collected judgements using CrowdFlower\footnote{\url{www.crowdflower.com}}, a crowdsourcing platform that distributes small tasks to a large number of workers, where we limited workers by their language expertise.
We note that while we elected to perform this additional stage of crowdsourcing, other researchers may find a fully automatic pipeline more desirable, so in our public release, we also release the pre-crowdsourced version of our MVSO.

\begin{table}[t]
  \resizebox{\columnwidth}{!}{
  \begin{tabular}{@{}l|c|c|c|c|c@{}}
     & \textbf{\#cand} & \textbf{\#users} & \textbf{\#coun} & \textbf{\%correct} & \textbf{\%agree} \\ \hline
    Arabic  &   81 &  10 &  7 & 0.57 & 0.90 \\
    Chinese & 1055 &  56 & 24 & 0.63 & 0.83 \\
    Dutch   & 1874 &  45 &  2 & 0.23 & 0.92 \\
    English & 5369 & 223 & 52 & 0.78 & 0.84 \\
    French  & 5840 & 152 & 37 & 0.43 & 0.86 \\
    German  & 3360 & 119 & 27 & 0.32 & 0.90 \\
    Italian & 4996 & 216 & 42 & 0.57 & 0.88 \\
    Persian &   65 &   6 &  6 & 0.37 & 0.86 \\
    Polish  &  159 &   6 &  1 & 0.52 & 0.93 \\
    Russian &  294 &  13 &  3 & 0.70 & 0.89 \\
    Spanish & 4992 & 190 & 30 & 0.70 & 0.89 \\
    Turkish &  701 &  61 & 22 & 0.66 & 0.84 
  \end{tabular}
  }
  \caption{Crowdsourcing results via no.\ of input candidate ANPs \#cand, \#users, countries \#coun, and perc.\ of ANPs accepted \%correct and annotator agreement \%agree.}
  \label{tab:anp_selection}
\end{table}

\subsubsection{Crowdsourcing Setup} \label{sec:experiment}
We required that each ANP was evaluated by at least three independent workers.
To ensure high quality results, we also required workers to be
(1) native speakers of the language, for which CrowdFlower had its own language competency and expertise test for workers, and
(2) have a good reputation according to the crowdsourcing platform, measured by workers' performance on other annotation jobs.
For whatever reasons, for three languages (Persian, Polish and Dutch), the CrowdFlower platform does not evaluate workers based on their language expertise, so we filtered them by provenience, selecting the countries according to the official language spoken (e.g.~Netherlands, Belgium, Aruba and Suriname for Dutch).

\textbf{Task Interface}:
The verification task for workers consisted of simply evaluating the correctness of adjective-noun pairs.
At the top of each page, we gave a short summary of the job and tasked workers: ``\textit{Verify that a word pair in} \texttt{<Language>} \textit{is a valid adjective-noun pair.}''
Workers were provided with a detailed definition of what an adjective-noun pair is and a summary of the criteria for evaluating ANPs, i.e.~it (1) is grammatically correct (adjective + noun), (2) shows language consistency, (3) shows generality, that is, commonly used and does not refer to a named entity, and (4) is semantically logical.
To guide workers, examples of correct and incorrect ANPs were provided for each criteria, where these ground truths were carefully judged and selected by four independent expert annotators.
In the interface, aside from instructions, workers were shown five ANPs and simply chose between ``yes'' or ``no'' to validate ANPs.

\textbf{Quality Control}:
Like some other crowdsourcing platforms, CrowdFlower provides a quality control mechanism called \textit{test questions} to evaluate and track the performance of workers.
These test questions come from pre-annotated ground truth, which in our case, correspond to ANPs with binary validation decisions for correctness.
To access our task at all, workers were first required to correctly answer at least seven out of ten such test questions.
In addition though, worker performance was tracked throughout the course of the task where these test questions were randomly inserted at certain points, disguised as normal units. 
For each language, we asked language experts to select ten correct and ten incorrect adjective-noun pairs from each language corpus to serve as the test questions.

\subsubsection{Crowdsourcing Results}
To measure the quality of our crowdsourcing, we looked at the annotator agreement along each validation task.
For all languages, the agreement was very strong with an average annotator agreement of 87\%, where workers agreed on either the correctness or incorrectness of ANPs.
We found that workers tended to agree more that ANPs were correct than that they were incorrect.
This was likely due to the wide range of possible criteria for rejecting an ANP where some criteria are easy to evaluate (e.g.~language consistency), while others, such as general usage versus named entity, may cause disagreement among users due to the cultural background of the worker.
For example, not all workers may agree that an ANP like \emph{big eyes} or \emph{big apple} refers to a named entity.
However, for languages where the agreement on the incorrect ANPs was high, namely Arabic, German, and Polish, the average annotator agreement as a percentage of all ANP for that language were greater than 90\%.

On average, our crowdsourcing validated that a vast number of the input candidate ANPs from our automatic ANP discovery and filtering process were indeed correct ANPs.
English, Spanish and Russian were the top three for which the automatic pipeline performed the best, where every three in five ANPs were approved by the crowd judgements.
However, for certain languages, including German, Dutch, Persian and French, the number of ANPs rejected by the crowd was actually greater than accepted ANPs due to a higher occurrence of mixed language pairs, e.g. \emph{witzig humor}.
In Table \ref{tab:anp_selection}, we summarize statistics from our crowdsourcing experiments according to the number of ANPs, percentage of correct/incorrect ANPs by worker majority vote, and average agreement.

%============================================================
\section{Dataset Analysis \& Statistics} \label{sec:analysis}

\begin{figure}[ht]
  \centering
  \includegraphics[scale=0.48]{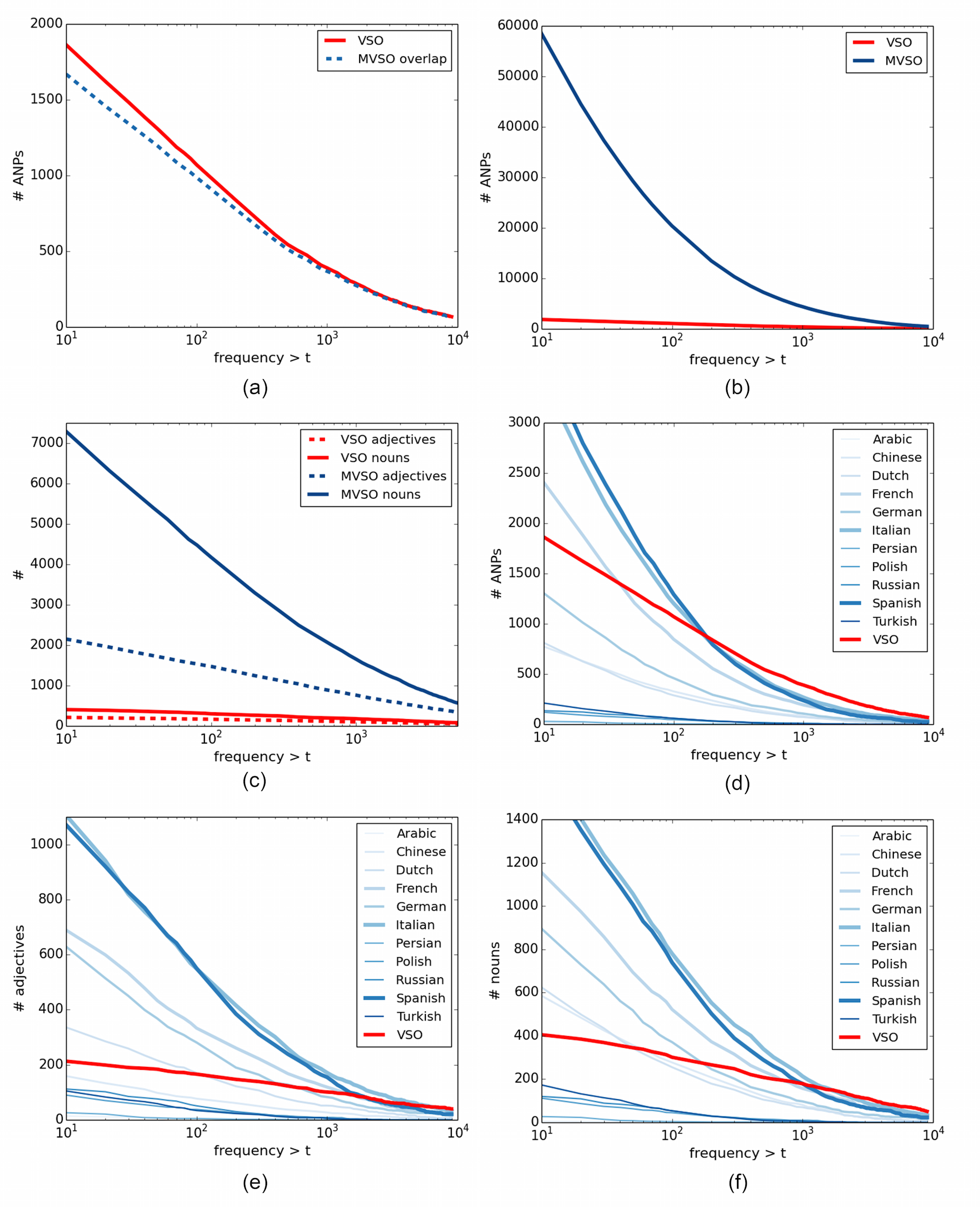}
  \caption{Comparison of our English MVSO and VSO \cite{borth_2013_vso} in Figures (a), (b) and (c), in terms of ANP overlap, no.\ of ANPs, adjectives and nouns; and with all other languages in Figures (d), (e) and (f), in terms of the no.\ of ANPs, adjectives and nouns when varying the frequency threshold $t$ from 0 to 10,000 (on log-scale), respectively.}
  \label{fig:comp}
\end{figure}

Having acquired a final set of adjective-noun pairs for each of the 12 languages, we downloaded images by querying the Flickr API with ANPs using a mix of tag and metadata search.
To limit the size of our dataset, we downloaded no more than 1,000 images per ANP query and also enforced a limit of no more than 20 images from any given uploader on Flickr for increased visual diversity.
The selected 1,000 images were selected from the pool of retrieved image tag search results, but in the event that this pool is less than 1,000, we also enlarged the pool to include searches on the image title and description, or metadata.
Selections from the pool of results were always randomized and a small number of images which Flickr or uploaders removed or changed privacy settings midway were removed.
In total, we downloaded 7,368,364 images across 15,630 ANPs for the 12 languages, where English (4,049,507), Spanish (1,417,781) and Italian (845,664) contributed the most images.

\subsection{Comparison with VSO \cite{borth_2013_vso}} \label{sec:comparison}

To verify and test the efficacy of our MVSO, we provide a comparison of our extracted English visual sentiment ontology with that of VSO \cite{borth_2013_vso} along dimensions of size (number of ANPs) and diversity of nouns and adjectives (Figure \ref{fig:comp}).
In Figure \ref{fig:comp}a, the overlap of English MVSO with VSO is compared with VSO alone after applying all filtering criteria except subsampling which might exclude ANPs belonging to VSO. 
As mentioned previously, about 86\% overlaps between them.
As we vary a frequency threshold $t$ (as described in Sec. \ref{sec:filtering}) over image tag counts, the overlap converges to 100\%.
This confirms that the popular ANPs covered by VSO are also covered by MVSO, an interesting finding given the difference in the crawling time periods and approaches.
In Figure \ref{fig:comp}b, we show that there are far greater number of ANPs in our English MVSO compared to VSO ANPs throughout all the possible values of frequency threshold, after applying all filtering criteria. Similarly, as shown in Figure \ref{fig:comp}c, given there are more adjectives and nouns in our English MVSO, we also achieve greater diversity than VSO. 

In Figure \ref{fig:comp}d, we compare the number of ANPs for the remaining languages in MVSO with VSO after applying all filtering criteria.
The curves show that VSO has more ANPs than all the languages for most of the languages over all values of $t$, except from Spanish, Italian and French in the low values of $t$.
Our intuition is that this is due to the popularity of English on Flickr compared to other languages.
In Figures \ref{fig:comp}e and \ref{fig:comp}d, we observe that these three languages have greater diversity of adjectives and nouns than VSO for $t \leq 10^3$, German and Dutch have greater diversity than VSO for smaller values of threshold $t$, while the rest of the languages have smaller diversity over most values of $t$. 

\subsection{Sentiment Distributions}
\begin{figure}[ht]
  \centering
  \includegraphics[scale=0.28]{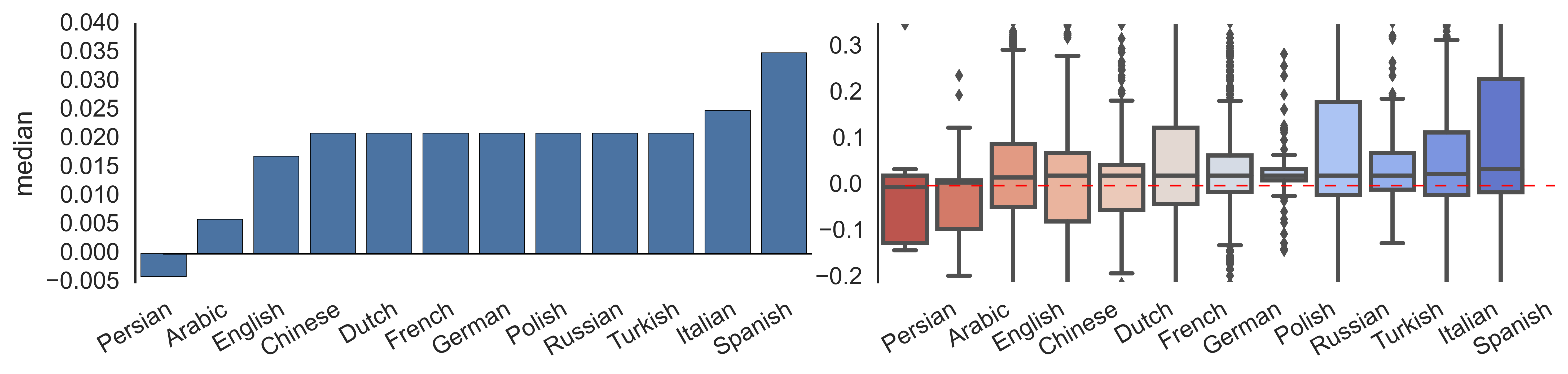}
  \caption{Median sentiment computed over all ANPs per language is shown on left, and the sentiment distribution using box plots on the right (zoomed at 90\% of the distributions).
On right, languages are sorted by median sentiment in ascending order (from the left).
}
  \label{fig:sent_dist}
\end{figure}

Returning to our research motivation from the Introduction, an interesting question to ask is which languages tend to be more positive or negative in their visual content.
To answer this, we computed the median sentiment value across all ANPs of each language and ranked languages as in Fig. \ref{fig:sent_dist}.
Here, to take into account the popularity difference among ANPs, we replicated each ANP $k$ times, with $k$ equal to the number of images tagged with the ANP, up to an upper limit $L=\alpha \times Avg_i$, where $Avg_i$ is the average image count per ANP in the $i$th language.
Varying $\alpha$ value will result in different medians and distributions, but the trend in differentiating positive languages from negative ones was quite stable.
We show the case when $\alpha = 3$ in Fig. \ref{fig:sent_dist}, indicating that there is an overall tendency toward positive sentiment across all languages, where Spanish demonstrates the highest positive sentiment, followed by Italian.
This surprising observation is in fact compatible with previous research showing that there is a universal positivity bias over languages with Spanish being the most relatively positive language \cite{dodds_2015}.
The languages with the lowest sentiment were Persian and Arabic, followed by English.

The sentiment distributions (Fig. \ref{fig:sent_dist}: right) also showed interesting phenomena:
Spanish being the most positive language also has the highest variation in sentiment, while German has the most concentrated sentiment distribution.
Even for languages that have the lowest median sentiment values, the range of sentiment was concentrated in a small range near zero (between 0 and -0.1).

\subsection{Emotion Distributions} \label{sec:emotionanalysis}
\begin{figure}[t]
  \centering
  \includegraphics[scale=0.33]{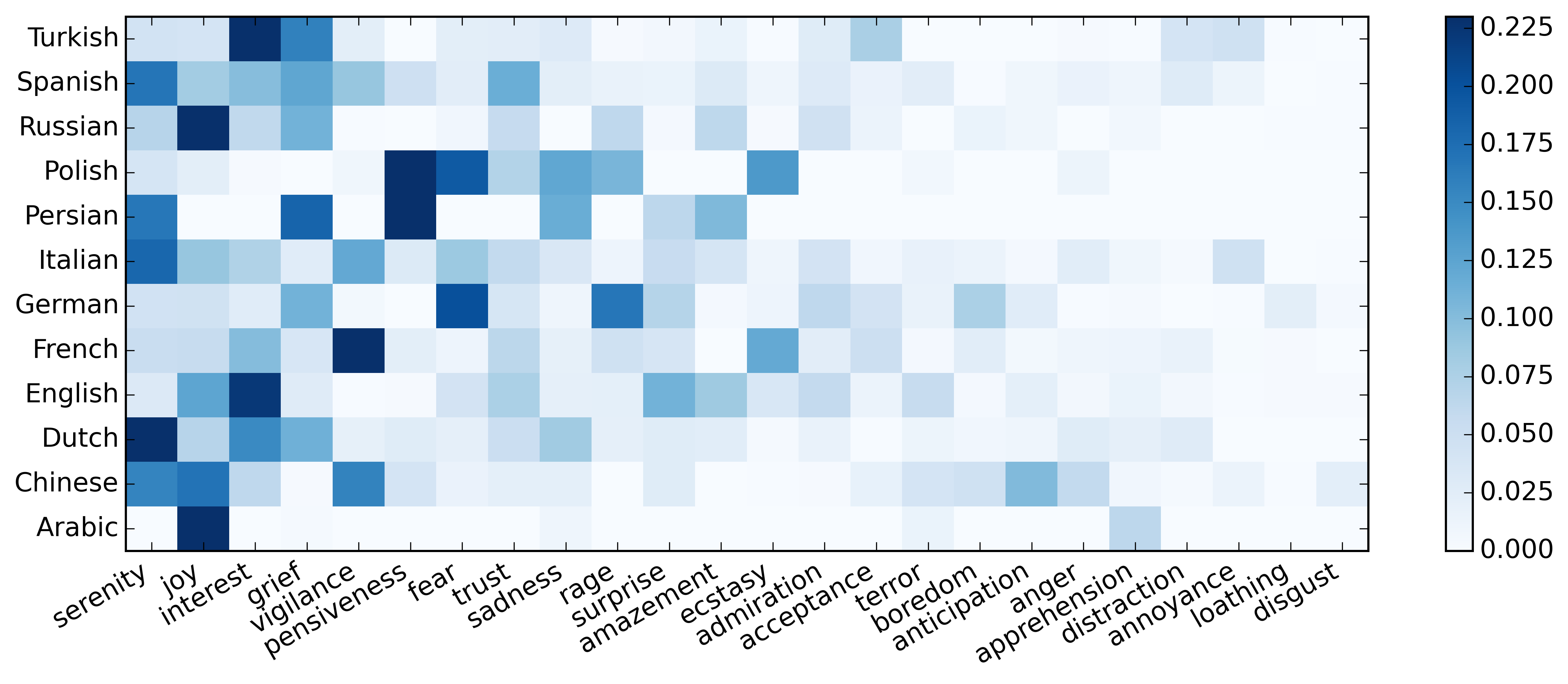}
  \caption{Probabilities of emotions per language with respect to their visual sentiment content. Emotions are ordered by the sum of their probabilities across languages (left to right) and clipped for better visualization. Each row sums to 1.}
  \label{fig:heatmap}
\end{figure} 

Another interesting question arises when considering co-occurrence of ANPs with the emotions in different languages.
While our adjective-noun pair concepts were selected to be sentiment-biased, emotions still represent the root of our framework since we built MVSO out from seed emotion terms.
So aside from sentiment, which focuses on only positivity/negativity, what are probable mappings of ANPs to emotions for each language?
What emotions are most frequently occurring across languages?
Given the set of keywords $E^{(l)}=\{e_{ij}^{(l)}\ |\ i=1\ldots 24, j=1\ldots n_{i} \}$ describing each emotion $i$ per language $l$, where $n_{i}$ is the number of keywords per emotion $i$, the set of ANPs belonging to language $l$, noted as $x \in X^{(l)}$, and the number of images tagged with both ANP $x$ and emotion keyword $e_{ij}$, $C^{(x)}=\{c_{ij}^{(x)}\ |\ i=1\ldots 24, j=1\ldots n_{i} \}$, we define the probabilities of emotion for each ANP $x$ in language $l$ as:
\begin{equation}
  \textrm{emo}^i(x) = \frac{\frac{1}{n_{i}} \sum_{j=1}^{n_{i}}c_{ij}^{(x)} }{ \sum_{i=1}^{24} \frac{1}{n_{i}} \sum_{j=1}^{n_{i}}c_{ij}^{(x)} } \in [0,1]
  \label{eq:emo}
\end{equation} 
Note the model in \eqref{eq:emo} does not take into account correlation among emotions, where for example, by an image tagged with ``ecstasy,'' users may also imply ``joy'' even though the latter is not explicitly tagged.
These correlations can be easily accounted for by smoothing co-occurrence counts $c_{ij}$ over correlated emotions, e.g.~the co-occurrence counts of an ANP tagged with ``ecstasy'' can be included partially in the co-occurrence count of ``joy.''
Regardless, still based on \eqref{eq:emo}, we compute a normalized emotion score per language $l$ and emotion $i$ as:
\begin{equation}
  \textrm{score}^i(l) = \frac{\sum_{x=1}^{|X^{(l)}|} \textrm{emo}^i(x) \cdot \textrm{count}(x) }{ \sum_{i=1}^{24} \sum_{x=1}^{|X^{(l)}|} \textrm{emo}^i(x) \cdot \textrm{count}(x) } \in [0,1]
  \label{eq:norm_emo}
\end{equation}
Figure \ref{fig:heatmap} shows these scores per language and Plutchik emotion \cite{plutchik_1980} on a heatmap diagram.
Scores in each row sum to 1 (over 24 emotions).
The emotions are ordered by the sum of their scores across languages.
The top-5 emotions across all languages are \emph{joy}, \emph{serenity}, \emph{interest}, \emph{grief} and \emph{fear}.
And the highest ranked emotion is \emph{joy} in Russian, Chinese and Arabic.
Two other emotions in the top-5 were also positive: \emph{serenity}, being high ranked emotion for Dutch, Italian, Chinese and Persian, and \emph{interest} for English, Turkish and Dutch.
The remaining two emotions in the top-5 were negative: \emph{grief} for Persian and Turkish, and \emph{fear}, which was high ranked in German and Polish.
We also observed that \emph{pensiveness} was top ranked for Persian and Polish, \emph{vigilance} for French, \emph{rage} for German, while \emph{apprehension} and \emph{distraction} for Spanish.
We note that these results are more concrete for languages with many ANPs (>1000) and less conclusive for those with few ANPs like Arabic and Persian.

%============================================================
\section{Cross-lingual Matching} \label{sec:anpmatch}
To get a gauge on the topics commonly mentioned across different cultures and languages, we analyzed alignments of translations for each ANP to English as a basis.
Two approaches were taken to study this: exact and approximate alignment.
We ensured that translations of ANPs also passed all our validation filters described in Sec. \ref{sec:filtering} for this analysis. 

\begin{figure}[t]
  \centering
  \includegraphics[scale=0.3]{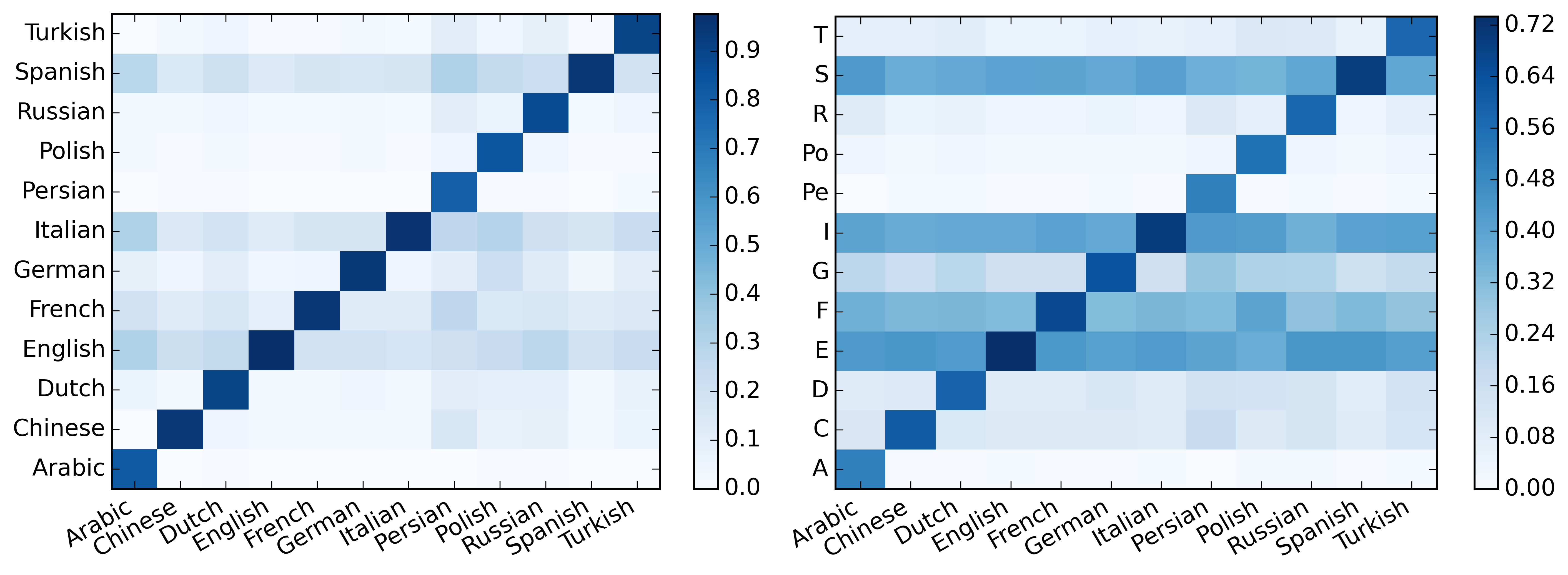}
  \caption{Percentage of times ANPs from one language (columns) were translated to [Left] the same phrase, or to [Right] the phrase in the same cluster as in another language (rows).}
  \label{fig:heatmap_joined}
\end{figure}

\textbf{Exact Alignment}: We grouped ANPs from each language that have the exact same translation.
For example, \emph{old books} was the translation for one or more ANPs from seven languages, including \begin{CJK}{UTF8}{bkai}老書\end{CJK} (Chinese), \emph{livres anciens} (French), \emph{vecchi libri} (Italian), \foreignlanguage{russian}{Старые книги} (Russian), \emph{libros antiguos} (Spanish), \emph{eski kitaplar} (Turkish).
The translation covered by the greatest number of languages was \emph{beautiful girl} with ANPs from ten languages.
Figure \ref{fig:heatmap_joined} (left) shows a correlation matrix of the number of times ANPs from pairs of languages appeared together in a set with the exact same translation, e.g.~out of all the translations that German ANPs were translated to (782), more of them were translated to the same phrase with the ANPs used by Dutch speakers (39) than with the ANPs used by Chinese speakers (23).
This was striking given that there were less (340) translation phrases from Dutch than from Chinese (473).

\begin{figure}[t]
  \centering
  \includegraphics[width=3.3in]{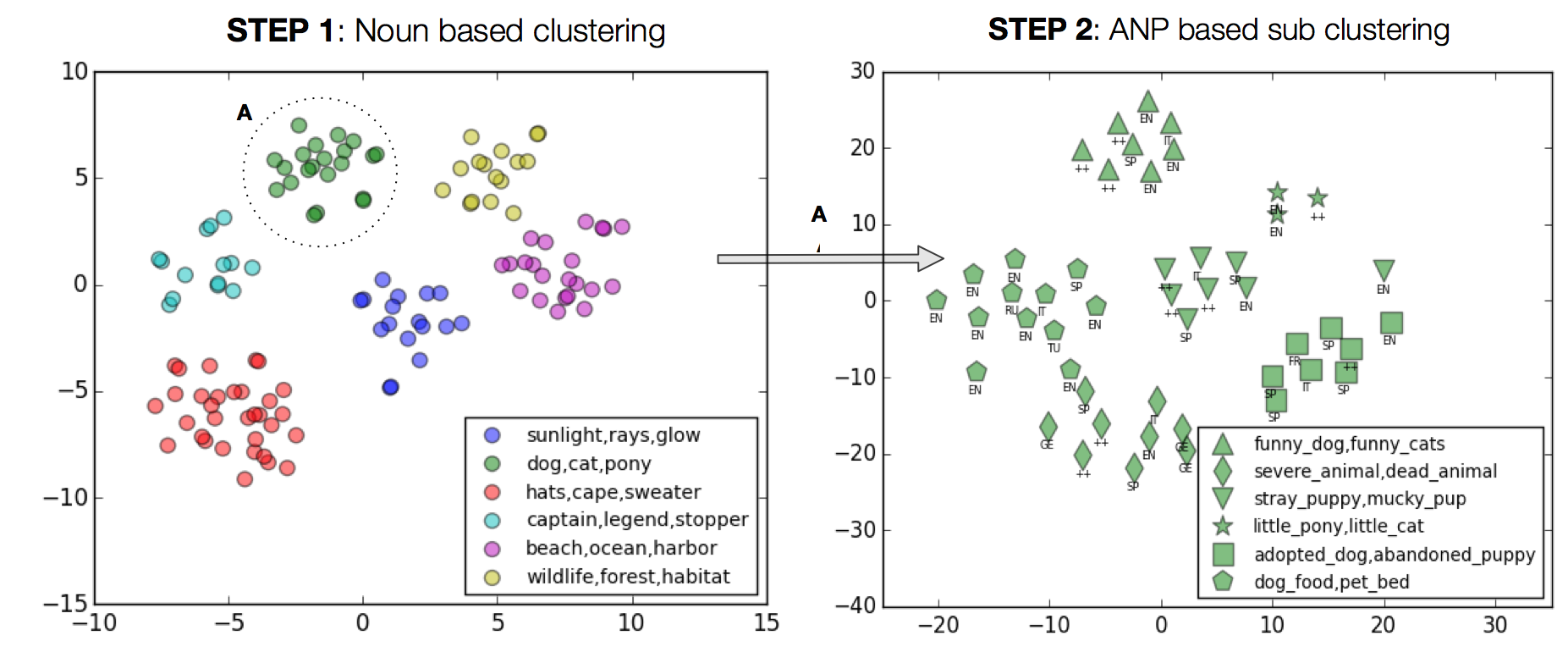}
  \caption{Examples of noun clusters (left) and ANP sub-clusters (right) from our two-stage clustering for cross-lingual matching. For visualization, word2vec \cite{mikolov_2013} vectors were projected to $\mathbb{R}^2$ using t-SNE \cite{maaten_2008}.}
  \label{fig:clustering_example}
\end{figure}

\textbf{Approximate Alignment}: Translations can be inaccurate, especially when capturing underlying semantics where context is not provided.
And so, we relaxed the strict condition for exact matches by approximately matching using a hierarchical two-stage clustering approach instead.
First, we extracted nouns using TreeTagger \cite{schmid_1994} from the list of translated phrases and discovered 3,099 total nouns.
We then extracted word2vec \cite{mikolov_2013} features, a word representation trained on a Google News\footnote{\url{news.google.com}} corpus, for these translated nouns (188 nouns were out-of-vocabulary), and performed $k$-means clustering ($k$=200) to get groups of nouns with similar meaning.
The number of clusters was picked based on the coherence of clusters; and we picked the number where the inertia value of the clustering started saturating while gradually increasing $k$. 
In the second stage of our hierarchical clustering, we split phrases from the translations into different groups based on the clusters their nouns belonged to.
We extracted word2vec \cite{mikolov_2013} features from the full translated phrase in each cluster and ran another round of $k$-means clustering (adjusting $k$ based on the number of phrases in each cluster, where phrases in each noun-cluster ranged from 3 to 253).
This two-stage clustering enables us to create a hierarchical organization of our ANPs across languages and form a multilingual ontology over visual sentiment concepts (MVSO), unlike the flat structure in VSO \cite{borth_2013_vso}.
We discovered 3,329 sub-clusters of ANP concepts, e.g.~resulting in clusters containing \emph{little pony} and \emph{little horse} as in Figure \ref{fig:clustering_example}.
This approach also yielded a larger intersection between languages, where German and Dutch share 118 clusters, and German and Chinese intersect over 101 ANP clusters.

The correlation matrix from this approximate matching is shown in Figure \ref{fig:heatmap_joined}, along with one subtree from our ontology by hierarchical clustering in Figure \ref{fig:clustering_example}.
For Figure \ref{fig:clustering_example}, we projected data to $\mathbb{R}^2$ using t-SNE dimensionality reduction \cite{maaten_2008}.
On the left, six clusters composed of different sets of nouns are shown with clusters of \textit{sunlight}-\textit{rays}-\textit{glow} and \textit{dog}-\textit{cat}-\textit{pony}.
On the right, we show the sub-clustering of ANPs for the \textit{dog}-\textit{cat}-\textit{pony} cluster in \textbf{A}, giving us noun groupings modified by sentiment-biasing adjectives to get ANPs like \textit{funny dog}-\textit{funny cats} and \textit{adopted dog}-\textit{abondoned puppy}.

%============================================================
\section{Visual Sentiment Prediction}
To test the effectiveness of a vision-based approach for visual affect understanding when crossing languages, we designed and built language-specific sentiment predictors using the data collected with MVSO.
Inspired by work in \cite{hu_2014}, we studied the extent to which the visual sentiments of a given language can be predicted by sentiment models of other languages.
We chose to focus on a sentiment prediction task, i.e.~predicting whether an image is of positive or negative sentiment, because there is a large body of work expressly focused on sentiment (e.g.~\cite{borth_2013_vso,yanulevskaya_2008,you_2014}) for its simplicity, compared to emotion prediction.
More importantly, we wanted to reduce the number of variables to be analyzed since our primary goal was to uncover cross-lingual differences.

We first constructed a bank of visual concept detectors like in \cite{borth_2013_sb} for our final MVSO adjective-noun pairs.
For simplicity, we focused on the six languages with the most ANPs and associated images in our dataset: in decreasing order, English, Spanish, Italian, French, German and Chinese.
Combined these six languages account for 94.7\% of the ANPs in MVSO and 98.4\% of the images in our dataset.
However, to ensure that there were enough training images for each ANP, only the ANPs with no less than 125 images were selected for model training and prediction.
This reduced the combined ANP coverage to 63.5\% but still ensured 92.0\% coverage for images.
For each ANP, the images were split randomly 80/20\% train/test, respectively.

\subsection{Visual Sentiment Concept Detectors} \label{sec:classifier}
To construct our bank of visual concept detectors of ANPs, we used convolutional neural networks (CNNs), in particular, adopting an AlexNet-styled architecture \cite{krizhevsky_2012} for its good performance on large-scale vision recognition and detection tasks.
To train our detector bank, we fined-tuned six models, one for each language, where network weights were initialized with DeepSentiBank \cite{chen_2014_dsb}, an AlexNet model trained on the VSO \cite{borth_2013_vso} dataset.
This fine-tuning approach ensures that each network begins with weights that are already somewhat ``affectively'' biased.
The base learning rates were set to 0.001 and the number of output neurons in the last fully connected layer were set to the number of training ANPs of each language.
Step sizes for reducing the learning rate in the second stage were set proportional to the number of training images per language.
For a single language, fine-tuning took between 12 and 40 hours for convergence on a single NVIDIA GTX 980 GPU implemented with Caffe \cite{jia_2014}.
From Table \ref{tab:classifiers}, as expected we achieve higher top-1 and top-5 accuracies than DeepSentiBank \cite{chen_2014_dsb}, even when the numbers of output neurons in English and Spanish are higher than those in \cite{chen_2014_dsb}.
Top-$k$ accuracy refers to the percentage of classifications that the true class is in the top $k$ predicted ranks.

\begin{table}[b]
  \resizebox{\columnwidth}{!}{
  \begin{tabular}{@{}lcccccrr@{}}
     & \textbf{\#ANPs} & \textbf{\#train} & \textbf{\#test} & \textbf{lrs (K)} & \textbf{time (hr)} & \textbf{top-1} & \textbf{top-5} \\ \hline
    English & 4,342 & 3,236,728 & 807,447 & 50 & 40 & 10.1\% & 21.7\% \\
    Spanish & 2,382 & 1,085,678 & 270,400 & 40 & 35 & 12.4\% & 25.4\% \\
    Italian & 1,561 &   602,424 & 149,901 & 30 & 30 & 17.0\% & 30.9\% \\
    French  & 1,115 &   462,522 & 115,112 & 30 & 26 & 17.7\% & 35.5\% \\
    German  &   275 &   108,744 &  27,048 & 20 & 12 & 30.1\% & 52.8\% \\
    Chinese &   243 &   102,740 &  25,575 & 20 & 15 & 27.1\% & 45.0\% \\
    DSB \cite{chen_2014_dsb} & 2,089 & 826,806 & 41,113 & - & - & 8.2\% & 19.1\% \\ \hline
  \end{tabular}
  }
  \caption{Adjective-noun pair (ANP) classification performance on Flickr images for six major languages in MVSO and compared to DeepSentiBank (DSB) \cite{chen_2014_dsb}. No.\ of visual sentiment concepts \#ANPs, \#train and \#test images along with learning rate step size (lrs, in thousands) are shown with training times (in hours), top-1 and top-5 accuracies.}
  \label{tab:classifiers}
\end{table}

\begin{figure}[t]
  \centering
  \includegraphics[scale=0.18]{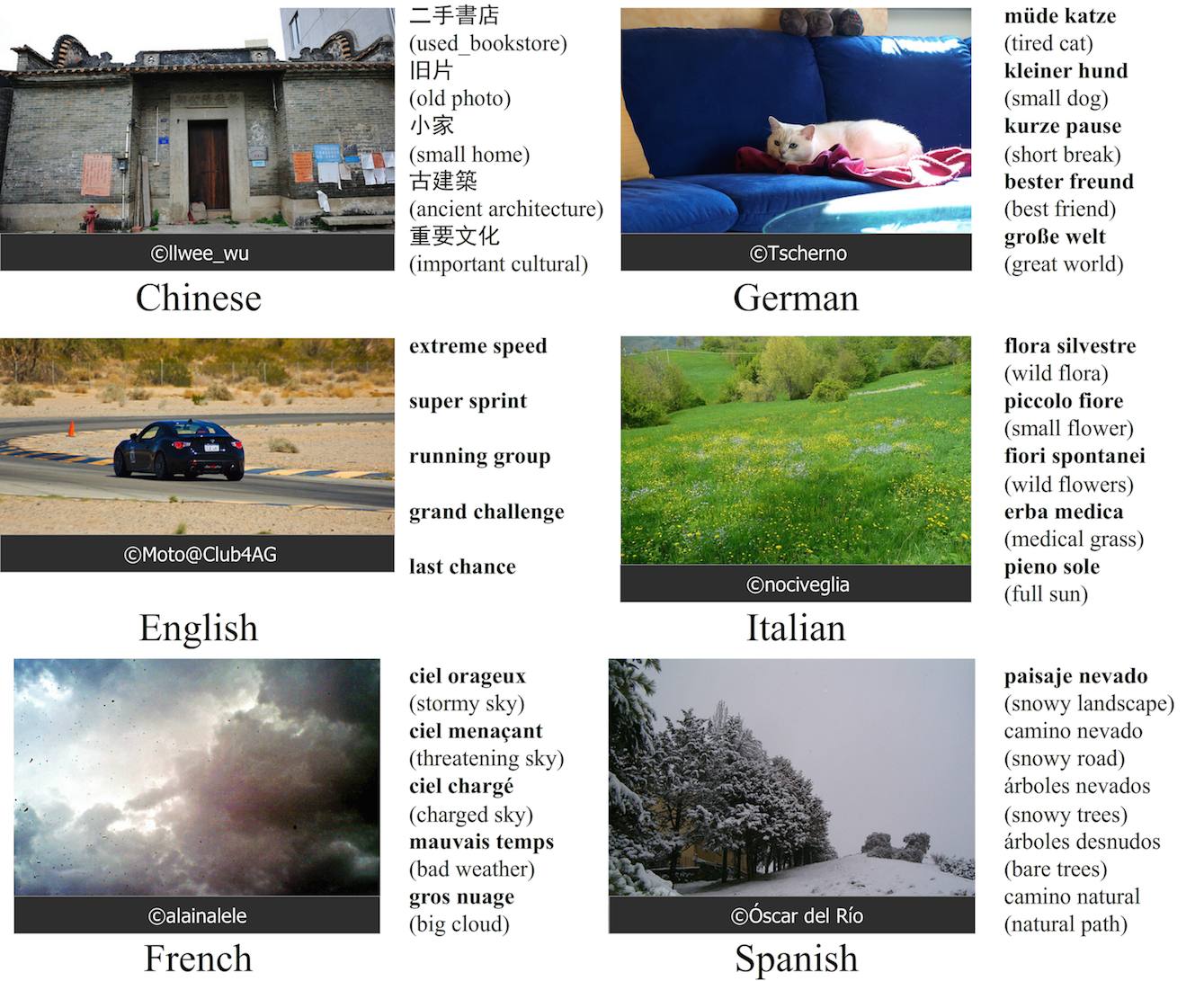}
  \caption{Example top-5 classification results from our multilingual visual sentiment detector bank. Translations to English provided for convenience.}
  \label{fig:detection_examples}
\end{figure}

\subsection{Sentiment Prediction on Flickr} \label{sec:sentiment}
We used the CNN-based visual concept models trained for each language to extract image features and use the sentiment scores of ANPs as supervised labels to learn sentiment prediction models.
We compared different layers of the CNN models as image features.
To simplify the process, we binarized the ANP sentiment scores computed with Eq.~\eqref{eq:sent}, i.e.~into positive and negative classes, and learned a binary classifier using linear SVMs, one for each language.
The training images are those associated with ANPs of strong sentiment scores (absolute values higher than 0.05).
Splits of training and test sets were stratified across all languages so that the amount of training and testing for positive and negative sentiment classes was the same for fair cross-lingual experiment comparison.
\begin{figure}[t]
  \centering
  \includegraphics[scale=0.38]{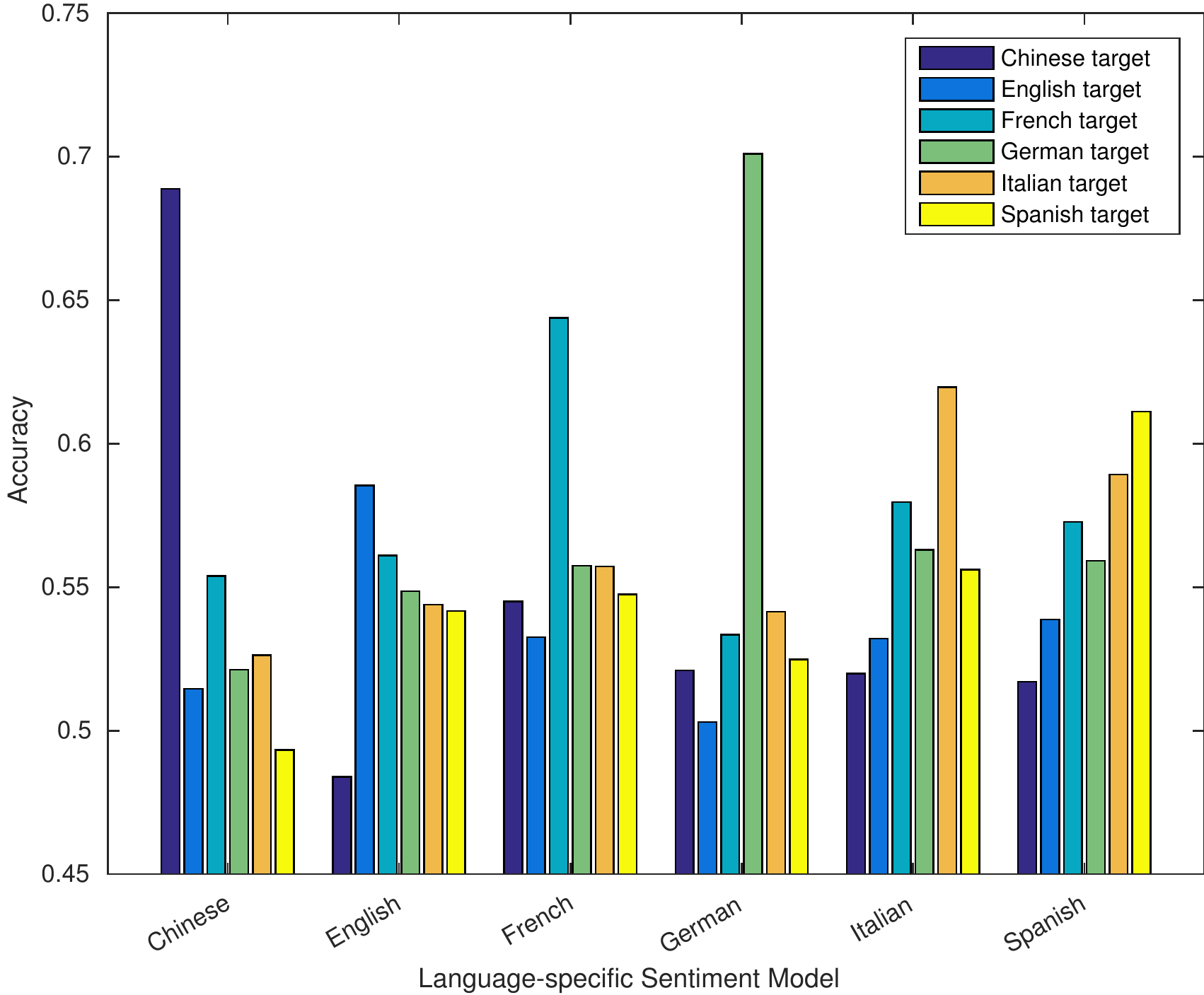}
  \caption{Image-based, cross-lingual domain transfer sentiment prediction results with language-specific models applied on cross-lingual examples.}
  \label{fig:sent_pred}
\end{figure}

We found that the softmax output features from the penultimate layer outputs of each language's CNN model performed the best for all languages, and we show resulting sentiment prediction results in Figure \ref{fig:sent_pred}.
Each language expectedly did better in predicting test samples from its own language, but in addition, Chinese generally was the most difficult to predict by models trained from other languages; and using a sentiment model trained over Chinese images to predict the sentiment in other languages was also the worst in average.
We speculate that this is due to the difference in the visual sentiment portrayal from Eastern and Western cultures.
Interestingly, the classification of French and Italian sentiments was the most consistent using models from all languages.
We also observed good performance in cross-lingual prediction for Latin languages, i.e.~Spanish, Italian and French, where Italian was the best cross-lingual classifier for Spanish and French sentiment, and Spanish was best for Italian sentiment, followed by French. 
Despite not performing as well as others in average, the English-specific sentiment model had the least variance in its accuracy across all languages, likely from the pervasiveness of English worldwide and across cultures.

\begin{figure}[h]
  \centering
  \includegraphics[width=3.3in]{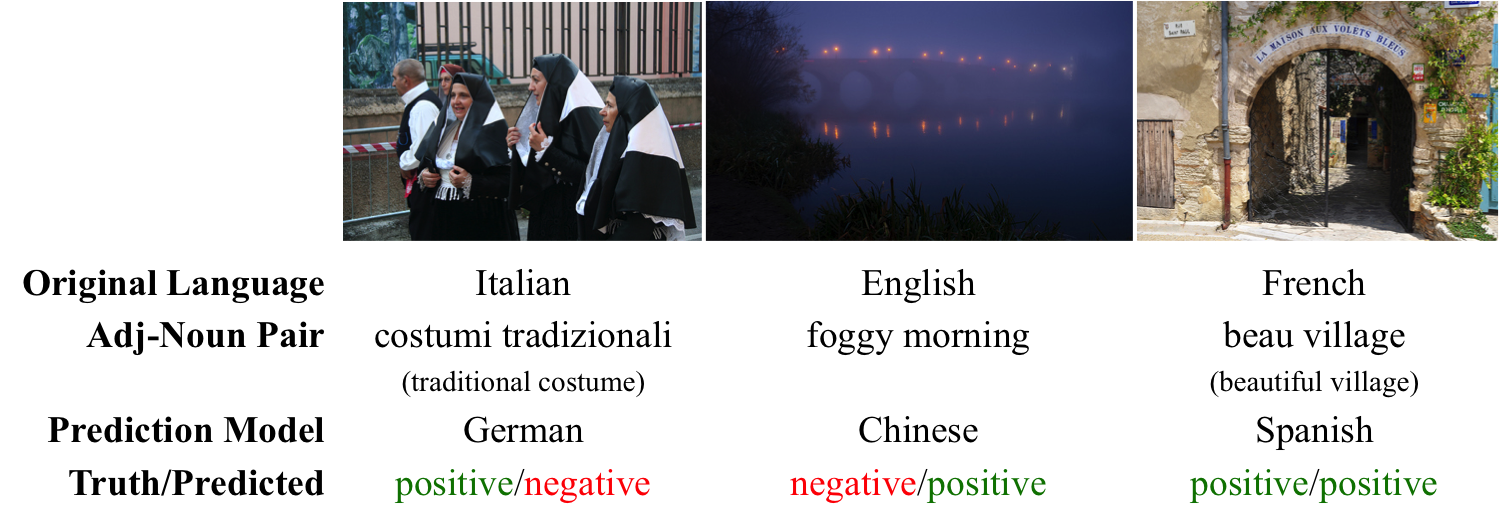}
  \caption{Classification examples from cross-lingual sentiment prediction. The model from a source language is used to predict the sentiment of a target language image where the true label comes from the sentiment of the associated ANP.}
  \label{fig:sent_ex}
\end{figure}

In Figure \ref{fig:sent_ex}, we show three classification example results from our cross-lingual sentiment prediction.
On the left, an image from the Italian test set representing the \emph{costumi tradizionali} concept was labeled as positive via sentiment scoring, but was predicted by the German model to be negative; this may be due to differences in cultural perceptions of traditional clothing.
In the center, the Chinese model wrongly predicted an image from the English test set of \emph{foggy morning} as positive, possibly for its resemblance to a Chinese painting.
And on the right, an image of a \emph{beau village} from the French test set was successfully classified as positive with the Spanish sentiment predictor.
These examples and preliminary experiments highlight some similarities and differences in how visual sentiment is expressed and perceived by various cultures.

%============================================================
\section{Conclusion \& Future Work} \label{sec:conclusion}
We proposed a new multilingual discovery method for visual sentiment concepts and showed its efficacy on a social multimedia platform for 12 languages.
We based our approach on the psychology theory that emotions are culture-specific and carry inherent linguistic context, and so we showed how to use language-specific part-of-speech labeling along with progressive filtering to achieve coverage and diversity of visual affect concepts in multiple languages.
In addition, we presented a two-stage hierarchical clustering approach to unify our ontology across languages.
And we make our Multilingual Visual Sentiment Ontology (MVSO), pre-crowdsourcing as well as post, and image dataset, available to the public.
A cross-lingual analysis of our large-scale MVSO and image dataset using semantic matching and visual sentiment prediction hint that emotions are not necessarily culturally universal.
Our preliminary results show that there are indeed commonalities, but also distinct separations, in how visual affect is expressed and perceived, where other works assumed only commonalities.
We believe these point to the colorful diversity of our world, rather than our inability to understand one another.

In the future, we plan to explore differences along other human factors which can be collected from self-reported user metadata like age group, gender, profession, etc.
We will also adopt our approach to other language-specific social multimedia platforms to counter the insufficient data for some languages like Arabic, Persian and Chinese.
In addition, while we discussed culture and languages in this work, we have not yet performed an in-depth study on geo-location data in MVSO, often provided along with uploaded images on Flickr.
While such information could be useful to distinguish between sub-cultures speaking the same languages (e.g.~Spanish vs.\ South-Americans), we omitted such a study here because of the noise that geo-location data can add.
For example, an American traveling in China uploading pictures is still more likely to use their native tongue to tag and sentimentally describe their content.
The trade-off is that while their semantics are culturally American, the uploaded visual content is now from another culture, so there is still much to be explored from geo-location and user metadata.

%============================================================
%
\section{Acknowledgments}

Research was sponsored by the U.S.~Defense Advanced Research Projects Agency (DARPA) under the Social Media in Strategic Communication (SMISC) program, Agreement Number W911NF-12-C-0028.
The views and conclusions contained in this document are those of the author(s) and should not be interpreted as representing the official policies, either expressed or implied, of the U.S.~Defense Advanced Research Projects Agency or the U.S.~Government.
The U.S.~Government is authorized to reproduce and distribute reprints for Government purposes notwithstanding any copyright notation hereon.
Co-author B.~Jou was supported by the Department of Defense (DoD) through the National Defense Science \& Engineering Graduate Fellowship (NDSEG) Program.
And co-author N.~Pappas was supported by the InEvent (FP7-ICT n.~287872) and MODERN Sinergia (CRSII2 147653) projects.

\bibliographystyle{abbrv}

\begin{thebibliography}{10}

\vspace{2mm}

\bibitem{balahur_2012}
A.~Balahur and M.~Turchi.
\newblock Multilingual sentiment analysis using machine translation?
\newblock In {\em WASSA}, 2012.

\bibitem{carmen_2008}
C.~Banea, R.~Mihalcea, J.~Wiebe, and S.~Hassan.
\newblock Multilingual subjectivity analysis using machine translation.
\newblock In {\em EMNLP}, 2008.

\bibitem{bautin_2008}
M.~Bautin, L.~Vijayarenu, and S.~Skiena.
\newblock International sentiment analysis for news and blogs.
\newblock In {\em ICWSM}, 2008.

\bibitem{borth_2013_sb}
D.~Borth, T.~Chen, R.~Ji, and S.-F. Chang.
\newblock Senti{B}ank: {L}arge-scale ontology and classifiers for detecting
  sentiment and emotions in visual content.
\newblock In {\em ACM MM}, 2013.

\bibitem{borth_2013_vso}
D.~Borth, R.~Ji, T.~Chen, T.~Breuel, and S.-F. Chang.
\newblock Large-scale visual sentiment ontology and detectors using adjective
  noun pairs.
\newblock In {\em ACM MM}, 2013.

\bibitem{chen_2014_dsb}
T.~Chen, D.~Borth, T.~Darrell, and S.-F. Chang.
\newblock Deep{S}enti{B}ank: {V}isual sentiment concept classification with
  deep convolutional neural networks.
\newblock {\em arXiv preprint arXiv:1410.8586}, 2014.

\bibitem{chen_2014}
Y.-Y. Chen, T.~Chen, W.~H. Hsu, H.-Y.~M. Liao, and S.-F. Chang.
\newblock Predicting viewer affective comments based on image content in social
  media.
\newblock In {\em ICMR}, 2014.

\bibitem{danglauser_2011}
E.~S. Dan-Glauser and K.~Scherer.
\newblock The {G}eneva affective picture database: {A} new 730-picture database
  focusing on valence and normative significance.
\newblock {\em Behav. Res. Meth.}, 43(2), 2011.

\bibitem{dodds_2015}
S.~Dodds and {et al.}
\newblock Human language reveals a universal positivity bias.
\newblock {\em PNAS}, 112(8), 2015.

\bibitem{wals}
M.~S. Dryer and M.~Haspelmath, editors.
\newblock {\em {WALS} Online}.
\newblock Max Planck Institute for Evolutionary Anthropology, 2013.
\newblock \url{http://wals.info/chapter/87}.

\bibitem{ekman_1993}
P.~Ekman.
\newblock Facial expression and emotion.
\newblock {\em American Psychologist}, 48(4), 1993.

\bibitem{esuli_2006}
A.~Esuli and F.~Sebastiani.
\newblock {SENTIWORDNET}: {A} publicly available lexical resource for opinion
  mining.
\newblock In {\em LREC}, 2006.

\bibitem{zelal_1994}
Z.~G\"{u}ng\"{o}rd\"{u} and K.~Oflazer.
\newblock Parsing {T}urkish using the lexical functional grammar formalism.
\newblock In {\em ACL}, 1994.

\bibitem{gygli_2013}
M.~Gygli, H.~Grabner, H.~Riemenschneider, F.~Nater, and L.~V. Gool.
\newblock The interestingness of images.
\newblock In {\em ICCV}, 2013.

\bibitem{halcasy_2007}
P.~Hal\'{a}csy, A.~Kornai, and C.~Oravecz.
\newblock Hun{P}os: An open source trigram tagger.
\newblock In {\em ACL}, 2007.

\bibitem{haselton_2006}
M.~G. Haselton and T.~Ketelaar.
\newblock Irrational emotions or emotional wisdom? {T}he evolutionary
  psychology of affect and social behavior.
\newblock {\em Affect in Soc. Think. and Behav.}, 8(21), 2006.

\bibitem{hu_2014}
X.~Hu and Y.-H. Yang.
\newblock Cross-cultural mood regression for music digital libraries.
\newblock In {\em JCDL}, 2014.

\bibitem{jia_2012}
J.~Jia, S.~Wu, X.~Wang, P.~Hu, L.~Cai, and J.~Tang.
\newblock Can we understand van {G}ogh's mood?: {L}earning to infer affects
  from images in social networks.
\newblock In {\em ACM MM}, 2012.

\bibitem{jia_2014}
Y.~Jia, E.~Shelhamer, J.~Donahue, S.~Karayev, J.~Long, R.~Girshick,
  S.~Guadarrama, and T.~Darrell.
\newblock Caffe: {C}onvolutional architecture for fast feature embedding.
\newblock In {\em ACM MM}, 2014.

\bibitem{jin_2010}
X.~Jin, A.~Gallagher, L.~Cao, J.~Luo, and J.~Han.
\newblock The wisdom of social multimedia: {U}sing {F}lickr for prediction and
  forecast.
\newblock In {\em ACM MM}, 2010.

\bibitem{jou_2014}
B.~Jou, S.~Bhattacharya, and S.-F. Chang.
\newblock Predicting viewer perceived emotions in animated {GIF}s.
\newblock In {\em ACM MM}, 2014.

\bibitem{ke_2006}
Y.~Ke, X.~Tang, and F.~Jing.
\newblock The design of high-level features for photo quality assessment.
\newblock In {\em CVPR}, 2006.

\bibitem{khosla_2014}
A.~Khosla, A.~Das~Sarma, and R.~Hamid.
\newblock What makes an image popular?
\newblock In {\em WWW}, 2014.

\bibitem{krizhevsky_2012}
A.~Krizhevsky, I.~Sutskever, and G.~E. Hinton.
\newblock Image{N}et classification with deep convolutional neural networks.
\newblock In {\em NIPS}, 2012.

\bibitem{lang_1997}
P.~Lang, M.~Bradley, and B.~Cuthbert.
\newblock International {A}ffective {P}icture {S}ystem ({IAPS}): {T}echnical
  manual and affective ratings.
\newblock Technical report, NIMH CSEA, 1997.

\bibitem{lee_2005}
J.~H. Lee, J.~S. Downie, and S.~J. Cunningham.
\newblock Challenges in cross-cultural/multilingual music information seeking.
\newblock In {\em ISMIR}, 2005.

\bibitem{machajdik_2010}
J.~Machajdik and A.~Hanbury.
\newblock Affective image classification using features inspired by psychology
  and art theory.
\newblock In {\em ACM MM}, 2010.

\bibitem{markus_1991}
H.~R. Markus and S.~Kitayama.
\newblock Culture and the self: {I}mplications for cognition, emotion, and
  motivation.
\newblock {\em Psychological Review}, 98(2), 1991.

\bibitem{mccarthy_1994}
E.~D. McCarthy.
\newblock The social construction of emotions: {N}ew directions from culture
  theory.
\newblock {\em Social Perspectives on Emotion}, 2, 1994.

\bibitem{mesquita_1997}
B.~Mesquita, N.~H. Frijda, and K.~Scherer.
\newblock Culture and emotion.
\newblock In J.~W. Berry, P.~R. Dasen, and T.~S. Saraswathi, editors, {\em
  Handbook of Cross-cultural Psychology}, volume~2. Allyn \& Bacon, 1997.

\bibitem{mihalcea_2007}
R.~Mihalcea, C.~Banea, and J.~Wiebe.
\newblock Learning multilingual subjective language via cross-lingual
  projections.
\newblock In {\em ACL}, 2007.

\bibitem{mikolov_2013}
T.~Mikolov, I.~Sutskever, K.~Chen, G.~Corrado, and J.~Dean.
\newblock Distributed representations of words and phrases and their
  compositionality.
\newblock In {\em NIPS}, 2013.

\bibitem{picard_1997}
R.~W. Picard.
\newblock {\em Affective Computing}.
\newblock MIT Press, 1997.

\bibitem{plutchik_1980}
R.~Plutchik.
\newblock {\em Emotion: {A} Psychoevolutionary Synthesis}.
\newblock Harper \& Row, 1980.

\bibitem{redi_2014}
M.~Redi, N.~O'Hare, R.~Schifanella, M.~Trevisiol, and A.~Jaimes.
\newblock 6 {S}econds of sound and vision: {C}reativity in micro-videos.
\newblock In {\em CVPR}, 2014.

\bibitem{russell_1991}
J.~A. Russell.
\newblock Culture and the categorization of emotions.
\newblock {\em Psychological Bulletin}, 110(3), 1991.

\bibitem{schmid_1994}
H.~Schmid.
\newblock Probabilistic part-of-speech tagging using decision trees.
\newblock In {\em Intl Conf. on New Methods in Language Proc.}, 1994.

\bibitem{thelwall_2010}
M.~Thelwall, K.~Buckley, G.~Paltoglou, and D.~Cai.
\newblock Sentiment strength detection in short informal text.
\newblock {\em Jour. Ameri. Soci. for Info. Sci. \& Tech.}, 61(12), 2010.

\bibitem{toutanova_2003}
K.~Toutanova, D.~Klein, C.~D. Manning, and Y.~Singer.
\newblock Feature-rich part-of-speech tagging with a cyclic dependency network.
\newblock In {\em NAACL}, 2003.

\bibitem{maaten_2008}
L.~van~der Maaten and G.~E. Hinton.
\newblock Visualizing high-dimensional data using t-{SNE}.
\newblock {\em JMLR}, 9, 2008.

\bibitem{vessel_2014}
E.~A. Vessel, J.~Stahl, N.~Maurer, A.~Denker, and G.~G. Starr.
\newblock Personalized visual aesthetics.
\newblock In {\em SPIE-IS\&T Electronic Imaging}, 2014.

\bibitem{yanulevskaya_2008}
V.~Yanulevskaya, J.~van Gemert, K.~Roth, A.~Herbold, N.~Sebe, and J.~M.
  Geusebroek.
\newblock Emotional valence categorization using holistic image features.
\newblock In {\em ICIP}, 2008.

\bibitem{you_2014}
Q.~You, J.~Luo, H.~Jin, and J.~Yang.
\newblock Robust image sentiment analysis using progressively trained and
  domain transferred deep networks.
\newblock In {\em AAAI}, 2014.

\bibitem{zajonc_1980}
R.~B. Zajonc.
\newblock Feeling and thinking: {P}references need no inferences.
\newblock {\em American Psychologist}, 35(2), 1980.

\end{thebibliography}

\end{document}